\documentclass[twocolumn,epjc3]{svjour3}

\usepackage{amsmath}
\usepackage{euscript,amssymb,epsfig}
\usepackage{graphicx}
\usepackage{amsfonts,latexsym}

\RequirePackage{graphicx}

\newcommand{\be}[1]{\begin{equation}\label{#1}}
\newcommand{\ee}{\end{equation}}
\newcommand{\ba}[1]{\begin{eqnarray}\label{#1}}
\newcommand{\ea}{\end{eqnarray}}
\newcommand{\rf}[1]{(\ref{#1})}
\newcommand{\nn}{\nonumber}

\journalname{Eur. Phys. J. C}

\begin{document}

\title{Scalar perturbations in cosmological models with quark nuggets}

\author{Maxim Brilenkov\thanksref{e1,addr1} \and Maxim Eingorn\thanksref{e2,addr2} \and Laszlo Jenkovszky\thanksref{e3,addr3} \and
Alexander Zhuk\thanksref{e4,addr4}}

\thankstext{e1}{e-mail: maxim.brilenkov@gmail.com}

\thankstext{e2}{e-mail: maxim.eingorn@gmail.com}

\thankstext{e3}{e-mail: jenk@bitp.kiev.ua}

\thankstext{e4}{e-mail: ai.zhuk2@gmail.com}

\institute{Department of Theoretical Physics, Odessa National University,\\ Dvoryanskaya st. 2, Odessa 65082, Ukraine\\ \label{addr1} \and North Carolina
Central University, CREST and NASA Research Centers,\\ Fayetteville st. 1801, Durham, North Carolina 27707, U.S.A.\\ \label{addr2} \and Bogolyubov Institute
for Theoretical Physics, Kiev 03680, Ukraine\\ \label{addr3} \and Astronomical Observatory, Odessa National University,\\ Dvoryanskaya st. 2, Odessa 65082,
Ukraine\\ \label{addr4}}

\date{Received: date / Accepted: date}

\maketitle

\begin{abstract}
In this paper we consider the Universe at the late stage of its evolution and deep inside the cell of uniformity. At these scales, the Universe is filled with
inhomogeneously distributed discrete structures (galaxies, groups and clusters of galaxies). Supposing that a small fraction of colored objects escaped
hadronization and survived up to now in the form of quark-gluon nuggets (QNs), and also taking into account radiation, we investigate scalar perturbations of
the FRW metrics due to inhomogeneities of dustlike matter as well as fluctuations of QNs and radiation. In particular, we demonstrate that the nonrelativistic
gravitational potential is defined by the distribution of inhomogeneities/fluctuations of both dustlike matter and QNs. Consequently, QNs can be distributed
around the baryonic inhomogeneities (e.g., galaxies) in such a way that it can solve the problem of the flatness of the rotation curves. We also show that the
fluctuations of radiation are caused by both the inhomogeneities in the form of galaxies and the fluctuations of quark-gluon nuggets. Therefore, if QNs exist,
the CMB anisotropy should contain also the contributions from QNs. Additionally, the spatial distribution of the radiation fluctuations is defined by the
gravitational potential. All these results look physically reasonable.
\end{abstract}

\keywords{quark gluon plasma \and scalar cosmological perturbations \and inhomogeneous Universe}

\section{Introduction}

It is well known that quark-gluon plasma can significantly affect the early dynamics of the Universe. For example, over two decades ago
\cite{LasloYad1,LasloYad2} (see also \cite{Tillmann}) the accelerated expansion of the early Universe was derived from a quark bag model with the proper
equations of state. It was called tepid \cite{LasloYad1,LasloYad2} or little \cite{Tillmann} inflation, in view of its moderate scales, compared to the better
known earlier inflation.

However, there is also a possibility that a small fraction of colored objects -- quarks and gluons -- escaped hadronization. They may survive as islands of
colored particles, called quark-gluon nuggets (for brevity sometimes also called quark nuggets (QNs)). This possibility was first considered by E. Witten
\cite{Witten} and scrutinized further in \cite{Applegate,Farhi,Chandra}. In his paper \cite{Witten}, Witten discusses the possibility that QNs can survive even
at zero temperature and pressure. If so, the "hot"\, quark-gluon phase in the form of QNs may affect the present expansion of the Universe. Indeed, in our
recent paper \cite{BEJZ1} we have shown that nuggets can contribute to dark matter provided that their interaction with ordinary matter is weak.

It is worth noting that the size distribution of QNs was calculated in \cite{Bhatta,Bhatta2}. The authors found that a large number of stable QNs exists in the
present Universe. They also claimed that QNs could be a viable candidate for cosmological dark matter. The survival probability of these QNs, i.e. the question
whether the primordial QNs can be stable on a cosmological time scale, is a key issue, and it was studied by a number of our predecessors. In particular, the
authors of \cite{BASR}, using the chromoelectric flux tube model, have demonstrated that QNs will survive against baryon evaporation if the baryon number of
the quark matter inside the nuggets is larger than $10^{42}$ which is a rather conservative estimate. A scenario where the Universe would be closed with QNs
with the baryon number density window $10^{39\div40}\leq N\leq 10^{49}$ or, in other words, the proverbial cosmological dark matter, containing $90\%$ or more
of all matter in the Universe, is made of QNs, was considered in the paper \cite{ARS}. The special role of the strange quark matter in the phase transition,
both in the context of the early Universe and in compact stars, was discussed in \cite{Gh}. A relativistic model for strange quark stars was proposed in
\cite{Kalam} (see also \cite{Zhitnitsky} for a different approach to get compact quark objects). Quark matter is believed to exist at the center of neutron
stars \cite{Perez}, in strange stars \cite{Drake} and as small pieces of strange matter \cite{Madsen}. The latter can result in ultra-high energy cosmic rays
\cite{Madsen2,Madsen3}. The search (in lunar soil and with an Earth orbiting magnetic spectrometer) for cosmic ray strangelets may be the most direct way of
testing the stable strange matter hypothesis.

In the present paper, we continue the investigation of the Universe filled with QNs. We consider the late stage of the Universe evolution when inhomogeneities
(such as galaxies and their groups) were already formed. Obviously, at this late and highly nonlinear stage the hydrodynamic approach is not adequate. Here,
the mechanical approach \cite{EZcosm1,EKZ2} is more appropriate. It works well inside the cell of uniformity \cite{EZcosm2}, and provides us a good tool to
investigate scalar perturbations for different cosmological models (see, e.g., \cite{BUZ1}). Therefore, it is of interest to study the compatibility of
cosmological models filled with nuggets with the mechanical approach. This is the main aim of our paper. As a result, we show that the considered models can be
compatible with the theory of scalar perturbations within the mechanical approach. It is worth noting that different variants of our model (more precisely, the
quark nugget model I and the quark-gluon plasma model I) were tested at cosmological scales using the experimental data from type Ia Supernovae, Long Gamma-Ray
Bursts and direct observations of the Hubble parameter in the recent paper \cite{Bilbao}. The authors found that, in general, these models do not contradict
the experimental data. We also demonstrate that the nonrelativistic gravitational potential is determined by the distribution of both the baryonic
inhomogeneities and QNs. Consequently, QNs can be distributed around the baryonic inhomogeneities (e.g., galaxies) in such a way that it can solve the problem
of the flatness of the rotation curves.

The paper is structured as follows. In Sec. 2, we briefly remind the background equations which describe the homogeneous and isotropic Friedmann cosmological
model with dustlike matter, radiation, quark-gluon nuggets and the cosmological constant. In Sec. 3, we investigate scalar perturbations of the FRW metrics.
Here, we demonstrate that QNs can be compatible with the theory of scalar perturbations. In Sec. 4, we find the QN distribution which allows the flat rotation
curves. The main results are briefly summarized in concluding Sec. 5.


\section{Background equations}

\setcounter{equation}{0}

In this section, we consider the homogeneous isotropic background cosmological model which satisfies Friedmann equations. As matter sources, we consider the
averaged dustlike matter (baryonic and dark matter{\footnote{As we mentioned in Introduction, QNs can play a role of dark matter. However, there is a
possibility of more than one type of dark matter. Therefore, we take into account also dustlike dark matter in our model.}}), radiation and quark nuggets. For
generality, we also include the cosmological constant.

\

{\em Quark-gluon nuggets}

\

The equation of state for quark-gluon plasma is not unique. There is a number of interesting modifications
\cite{LasloYad1,LasloYad2,Kallmann,LasloEch,Begun2004,Pisarski2006,Pisarski2007}. In our paper \cite{BEJZ1}, we considered two possible forms of the equation
of state. The corresponding total background pressure{\footnote{This is the summarized pressure inside of all nuggets averaged over the whole Universe. }} and
energy density of all nuggets in the Universe, as well as their temperature, read, respectively,
\ba{2.1} &{}&\bar{p}_{\mathrm{QN}}=\frac{A_1T+A_4T^4}{a^3}, \quad \bar{\varepsilon}_{\mathrm{QN}}=\frac{3A_4T^4}{a^3},\nn\\
&{}&T=\left(\frac{\left(C/a\right)^{3/4}-A_1}{A_4}\right)^{1/3} \ea
for Model I, and
\ba{2.2} &{}&\bar{p}_{\mathrm{QN}}=\frac{A_0+A_4T^4}{a^3}, \quad \bar{\varepsilon}_{\mathrm{QN}}=\frac{-A_0+3A_4T^4}{a^3},\nn\\
&{}&T=\left(\frac{\left(C/a\right)-A_0}{A_4}\right)^{1/4} \ea
for Model II. Here, $a$ is the scale factor of the Universe, $C$ is the constant of integration and parameters $A_0,A_1,A_4$ are defined by the bag model
constants and satisfy the relations \cite{BEJZ1}:
\be{2.3}
\frac{A_1}{A_4}= -0.8114 \; T_c^3\, ,\quad \frac{A_0}{A_4}= -0.8114 \; T_c^4\, ,
\ee
where $T_c \approx 200$ MeV. It is also worth noting that $A_0,A_1 <0$ and $A_4>0$.

\

{\em Friedmann equations}

\

For our models, the Friedmann equations read
\be{2.4} \frac{3\left(\mathcal{H}^2+\mathcal{K}\right)}{a^2}= \kappa\left(\overline{T}^0_0+\overline \varepsilon_{\mathrm{rad}} + \overline
\varepsilon_{\mathrm{QN}} \right) +\Lambda \ee
and
\be{2.5}
\frac{2\mathcal{H}'+\mathcal{H}^2+\mathcal{K}}{a^2}=-\kappa\left(\overline p_{\mathrm{rad}}+ \overline p_{\mathrm{QN}}\right) + \Lambda\, ,
\ee
where ${\mathcal H}\equiv a'/a\equiv (da/d\eta)/a$, \ $\kappa\equiv 8\pi G_N/c^4$ ($c$ is the speed of light and $G_N$ is the Newton's gravitational constant)
and $\mathcal K=-1,0,+1$ for open, flat and closed Universes, respectively. Conformal time $\eta$ and synchronous time $t$ are connected as $cdt=a d\eta$.
Here, $\overline T^{i}_k$ is the energy-momentum tensor of the average pressureless dustlike matter. For such matter, the energy density $\overline T^{0}_{0}
=\overline \rho c^2/a^3$ is the only nonzero component. $\overline \rho=\mbox{const}$ is the comoving average rest mass density \cite{EZcosm1}. As usual, for
radiation we have the equation of state: $\overline p_{\mathrm{rad}}=(1/3)\overline \varepsilon_{\mathrm{rad}}$. From Eqs. \rf{2.4} and \rf{2.5}, we can easily
get the following auxiliary equation:
\ba{2.6} &{}&\frac{2}{a^2}\left(\mathcal{H}'-\mathcal{H}^2-\mathcal{K}\right)\nn\\
&=& -\kappa\left(\overline{T}^0_0+\overline \varepsilon_{\mathrm{rad}} +\overline \varepsilon_{\mathrm{QN}} + \overline p_{\mathrm{rad}}+ \overline
p_{\mathrm{QN}}\right)\, . \ea


\section{Scalar perturbations}

\setcounter{equation}{0}

As we have written in Introduction, we consider the Universe at late stages of its evolution when galaxies and clusters of galaxies have already formed. At
scales much larger than the characteristic distance between these inhomogeneities, the Universe is well described by the homogeneous and isotropic FRW metrics.
This is approximately 190 Mpc and larger \cite{EZcosm2}. At these scales, the matter fields (e.g., cold dark matter) are well described by the hydrodynamical
approach.  However, at smaller scales the Universe is highly inhomogeneous. Here, the mechanical approach looks more adequate \cite{EZcosm1,EZcosm2}.

In the mechanical approach, galaxies, dwarf galaxies and clusters of galaxies (composed of baryonic and dark matter) can be considered as separate compact
objects. Moreover, at distances much greater than their characteristic sizes they can be well described as point-like matter sources. This is generalization of
the well-known astrophysical approach \cite{Landau} (see \S 106) to the case of dynamical cosmological background. Usually, the gravitational fields of these
inhomogeneities are weak and their peculiar velocities are much less than the speed of light. Therefore, we can construct a theory of perturbations where the
considered point-like inhomogeneities perturb the FRW metrics. Quark-gluon nuggets and radiation can also fluctuate. All these fluctuations result in scalar
perturbations of the FRW metrics. In the conformal Newtonian gauge, such perturbed metrics is \cite{Mukhanov,Rubakov}
\be{3.1}
ds^2\approx a^2\left[(1+2\Phi)d\eta^2-(1-2\Psi)\gamma_{\alpha\beta}dx^{\alpha}dx^{\beta}\right]\, ,
\ee
where scalar perturbations $\Phi,\Psi \ll 1$. Following the standard argumentation, we can put $\Phi=\Psi$. We consider the Universe at the late stage of its
evolution when the peculiar velocities of inhomogeneities/fluctuations are much less than the speed of light:
\be{3.2}
\frac{dx^{\alpha}}{d\eta}
=a\frac{dx^{\alpha}}{dt} \frac{1}{c}\equiv \frac{v^{\alpha}}{c}\ll 1\, .
\ee
We should stress that smallness of the nonrelativistic gravitational potential $\Phi$ and smallness of peculiar velocities $v^{\alpha}$ are two independent
conditions (e.g., for very light relativistic masses the gravitational potential can still remain small). Under these conditions, the gravitational potential
$\Phi$ satisfies the following system of equations (see \cite{EZcosm1,EZcosm2} for details):
\ba{3.3} &{}&\Delta\Phi-3\mathcal{H}(\Phi'+\mathcal{H}\Phi)+3\mathcal{K}\Phi\nn\\
&=&
\frac{1}{2}\kappa a^2\left(\delta T_0^0+\delta\varepsilon_{\mathrm{QN}}+\delta\varepsilon_{\mathrm{rad1}}+\delta\varepsilon_{\mathrm{rad2}}\right)\, ,\\
\label{3.4}
&{}&\frac{\partial}{\partial x^\beta}(\Phi'+\mathcal{H}\Phi)=0\, ,\\
\label{3.5} &{}&\Phi''+3\mathcal{H}\Phi'+(2\mathcal{H}'+\mathcal{H}^2)\Phi-\mathcal{K}\Phi\nn\\
&=& \frac{1}{2}\kappa a^2\left(\delta p_{\mathrm{QN}}+\delta p_{\mathrm{rad1}}+\delta p_{\mathrm{rad2}}\right)\, , \ea
where the Laplace operator $\triangle$ is defined with respect to the metrics $\gamma_{\alpha\beta}$.

Following the reasoning of \cite{EZcosm1,EZcosm2}, we took into account that peculiar velocities of inhomogeneities are nonrelativistic, and under the
corresponding condition \rf{3.2} the contribution of $\delta T_{\beta}^0$ is negligible compared to that of $\delta T_{0}^0$ both for dustlike matter and the
considered quark-gluon nuggets and radiation{\footnote{For all considered matter sources, the nondiagonal components of the energy-momentum tensor $\delta
T_{\beta}^0$ are connected with the peculiar velocities of their inhomogeneities/fluctuations (see the corresponding discussion in \cite{BUZ1}).}}. In other
words, account of $\delta T_{\beta}^0$ is beyond the accuracy of the model. This approach is completely consistent with \cite{Landau} where it is shown that
the nonrelativistic gravitational potential is defined by the positions of the inhomogeneities but not by their velocities (see Eq. (106.11) in this book).

From Eq. \rf{3.4} we get
\be{3.6}
\Phi(\eta,{\bf r})=\frac{\varphi({\bf r})}{c^2a(\eta)}\, ,
\ee
where $\varphi({\bf r})$ is a function of all spatial comoving coordinates and we have introduced $c^2$ in the denominator for convenience. In the vicinity of
an inhomogeneity, the comoving potential $\varphi({\bf r})\sim 1/r$ \cite{EZcosm1,EZcosm2,BUZ1}, and the nonrelativistic gravitational potential
$\Phi(\eta,{\bf r})\sim 1/(a r)=1/R$, where $R=ar$ is the physical distance. Hence, $\Phi$ has the correct Newtonian limit near the inhomogeneities.

In \rf{3.3} $\delta T^0_0$ is related to the fluctuation of the energy
density of dustlike matter and has the form \cite{EZcosm1}:
\be{3.7}
\delta T_{0}^0=\frac{\delta\rho c^2}{a^3}+\frac{3\overline{\rho}c^2\Phi}{a^3}\, ,
\ee
where $\delta\rho$ is the difference between the real and average rest mass densities: $\delta\rho = \rho-\overline\rho$.

In Eqs. \rf{3.3} and \rf{3.5}, we split the fluctuations of radiation into two parts. Here, the part labeled by "rad1" is caused by the inhomogeneities of
dustlike matter (e.g., by galaxies and their groups), and the part labeled by "rad2" is related to fluctuations of quark-gluon nuggets. For both of them, we
have the same equations of state: $\delta p_{\mathrm{rad1}}=(1/3)\delta\varepsilon_{\mathrm{rad1}}$ and $\delta
p_{\mathrm{rad2}}=(1/3)\delta\varepsilon_{\mathrm{rad2}}$. We have shown in \cite{EZcosm2} that $\delta\varepsilon_{\mathrm{rad1}}$ has the form:
\be{3.8}
\delta\varepsilon_{\mathrm{rad1}}=-\frac{3\overline{\rho}\varphi}{a^4}\, .
\ee

Taking into account Eqs. \rf{2.6}, \rf{3.6}, \rf{3.7} and \rf{3.8}, we can rewrite Eqs. \rf{3.5} and \rf{3.3} as follows:
\ba{3.9}
-\left(\overline{\varepsilon}_{\mathrm{QN}}+\overline{p}_{\mathrm{QN}}\right)\frac{\varphi}{c^2a}&=&\delta p_{\mathrm{QN}}+
\frac{1}{3}\delta\varepsilon_{\mathrm{rad2}}\, , \\
\label{3.10} \triangle\varphi+3\mathcal{K}\varphi=\frac{\kappa c^4}{2}\delta\rho&+&\frac{\kappa c^2 a^3}{2}\delta\varepsilon_{\mathrm{QN}}+\frac{\kappa
c^2a^3}{2}\delta\varepsilon_{\mathrm{rad2}}\, .\nn\\ \ea
To get Eq. \rf{3.9}, we took into account that we consider this equation up to terms $O(1/a^4)$ inclusive. The terms $\overline \varepsilon_{\mathrm{rad}}
\Phi,\ \overline p_{\mathrm{rad}} \Phi \sim O(1/a^5)$ and we dropped them. Let us investigate the system of equations \rf{3.9} and \rf{3.10} separately for
Model I and Model II.

\

{\em Model I}

\

We consider first Eq. \rf{3.9}. As noted above, we keep in this equation terms up to $O(1/a^4)$. Therefore, the sum
$\overline{\varepsilon}_{\mathrm{QN}}+\overline{p}_{\mathrm{QN}}$ should not include terms of the order of smallness higher than $1/a^3$. It is useful to
introduce an auxiliary quantity $\xi \equiv (C/a)^{3/4}$. Then, using the formulae \rf{2.1}, this sum takes the form
\ba{3.11}
\overline{\varepsilon}_{\mathrm{QN}}+\overline{p}_{\mathrm{QN}} &=& \frac{1}{a^3A^{1/3}_4}\left\{3\left[\xi-A_1\right]^{4/3}+\xi
\left[\xi-A_1\right]^{1/3}\right\}\nn\\
&=&\frac{3(-A_1)^{4/3}}{a^3A^{1/3}_4}+\frac{5(-A_1)^{1/3}}{a^3A^{1/3}_4}\xi+\frac{1}{a^3}o(\xi)\nn\\
&\approx& \frac{3(-A_1)^{4/3}}{a^3A^{1/3}_4}\, . \ea
Similarly, on the right hand side of \rf{3.9} $\delta p_{\mathrm{QN}}$ also should not contain the terms of the order of smallness higher than $1/a^4$.
Obviously, the same should hold for $\delta \varepsilon_{\mathrm{QN}}$.

Now, we need to make the important remark. We suppose that fluctuations of quark-gluon nuggets are caused by two reasons. First, it is the fluctuation of the
distribution of QNs (i.e. the fluctuation of the number density of QNs). We will define it by a new function $f(\bf{r})$. Second, it is the fluctuations of the
temperature of QNs $\delta T$. Therefore, from formulae \rf{2.1}, we have
\ba{3.12}
\delta\varepsilon_{\mathrm{QN}}&=&\frac{3A_4T^4}{a^3}f({\bf r})+\frac{12A_4T^3}{a^3}\delta T\, ,\\
\label{3.13}
\delta p_{\mathrm{QN}}&=&\frac{A_1T+A_4T^4}{a^3}f({\bf r}) + \frac{A_1+4A_4T^3}{a^3}\delta T\, .
\ea
Then, we get
\ba{3.14}
\delta\varepsilon_{\mathrm{QN}}&=&
\left[\frac{A_1}{a^3}\left(\frac{\xi-A_1}{A_4}\right)^{1/3}+\frac{A_4}{a^3} \left(\frac{\xi-A_1}{A_4}\right)^{4/3}\right]f({\bf
r})\nn\\
&+&\left[\frac{A_1}{a^3}+\frac{4A_4}{a^3}\left(\frac{\xi-A_1}{A_4}\right)\right]\delta T \nn \\
&\approx& \frac{3A_4}{a^3}\left[\left(\frac{-A_1}{A_4}\right)^{4/3}f({\bf r})-\frac{4A_1}{A_4}\delta T\right]\nn\\
&+&\frac{4}{a^3}\left[\left(\frac{-A_1}{A_4}\right)^{1/3}f({\bf r})+3\delta T\right]\left(\frac{C}{a}\right)^{3/4}\, \ea
and
\ba{3.15} \delta p_{\mathrm{QN}}&\approx& -\frac{3A_1}{a^3}\delta T\nn\\
&+&\frac{1}{a^3}\left[\left(\frac{-A_1}{A_4}\right)^{1/3}f({\bf r})+4\delta T\right]\left(\frac{C}{a}\right)^{3/4}\, . \ea
Hence, Eq. \rf{3.9} reads
\ba{3.16} -\frac{3(-A_1)^{4/3}}{a^4A^{1/3}_4}\frac{\varphi}{c^2}&=&-\frac{3A_1}{a^3}\delta T\nn\\
&+&\left[\left(\frac{-A_1}{A_4}\right)^{1/3}f({\bf r})+4\delta T\right]\left(\frac{C}{a^5}\right)^{3/4}\nn\\
&+&\frac{1}{3}\delta\varepsilon_{\mathrm{rad2}}\, . \ea
We can use this equation to determine the fluctuations of the temperature of QNs:
\ba{3.17} \delta T &\approx& \frac{1}{3A_1}\left[\frac{3(-A_1)^{4/3}}{A^{1/3}_4}\frac{\varphi}{ac^2}
+\frac{1}{3}a^3\delta\varepsilon_{\mathrm{rad2}}\right.\nn\\
&+&\left.\left(\frac{-A_1}{A_4}\right)^{1/3}\left(\frac{C}{a}\right)^{3/4}f({\bf r})\right]\, . \ea

Let us turn now to Eq. \rf{3.10}. Taking into account relations  \rf{3.14} and \rf{3.17}, we can write it as follows:
\ba{3.18} \triangle\varphi+3\mathcal{K}\varphi&\approx& \frac{\kappa c^4}{2}\delta\rho+\frac{\kappa c^2a^3}{2}\delta\varepsilon_{\mathrm{rad2}}\nn\\
&+& \frac{3\kappa c^2A_4}{2}\left[\left(\frac{-A_1}{A_4}\right)^{4/3}f({\bf r})-\frac{4A_1}{A_4}\delta T\right]\nn \\
&+&2 \kappa c^2\left[\left(\frac{-A_1}{A_4}\right)^{1/3}f({\bf r})+3\delta T\right]\left(\frac{C}{a}\right)^{3/4}
\nn \\
&\approx& \frac{\kappa c^4}{2}\delta\rho +\frac{3\kappa c^2}{2}\frac{(-A_1)^{4/3}}{(A_4)^{1/3}}f({\bf r})\nn\\
&-&\frac{6(-A_1)^{4/3}\kappa}{aA^{1/3}_4}\varphi -\frac{\kappa c^2a^3}{6}\delta\varepsilon_{\mathrm{rad2}}\, . \ea
Therefore, we arrive at the system of two equations:
\be{3.19}
\triangle\varphi+3\mathcal{K}\varphi=\frac{\kappa c^4}{2}\delta\rho+\frac{3\kappa c^2}{2}\frac{(-A_1)^{4/3}}{(A_4)^{1/3}}f({\bf r})\, ,
\ee
and
\ba{3.20} -\frac{6(-A_1)^{4/3}\kappa}{aA^{1/3}_4}\varphi&-&\frac{\kappa c^2a^3}{6}\delta\varepsilon_{\mathrm{rad2}}=0 \quad \Rightarrow \quad \nn\\
\delta\varepsilon_{\mathrm{rad2}}&=&-\frac{36(-A_1)^{4/3}}{c^2a^4A^{1/3}_4}\varphi\, . \ea
Eq. \rf{3.19} demonstrates that the gravitational potential is defined by the functions of fluctuation distribution of dustlike matter $\delta \rho (\bf r)$
and quark-gluon nuggets $f(\bf r)$. Eq. \rf{3.20} shows that, similar to \rf{3.8}, $\delta\varepsilon_{\mathrm{rad2}}\sim 1/a^4$ which is the physically
reasonable result for fluctuations associated with radiation. Additionally, the spatial distribution of these fluctuations is defined by the gravitational
potential $\varphi (\bf r)$ (similar to \rf{3.8}) that is also reasonable.

\

{\em Model II}

\

Now, we consider the Model II which is defined by the background equations \rf{2.2}. The procedure is absolutely similar to the calculations carried out for
the Model I. As a result, for the QN temperature fluctuations we get
\ba{3.21} \delta T &\approx& \frac{1}{A_0}\left(\frac{-A_0}{A_4}\right)^{1/4}\left[-A_0\frac{\varphi}{ac^2}\right.\nn\\
&+&\left.\frac{1}{12}a^3\delta\varepsilon_{\mathrm{rad2}} +\frac{C}{4a}f({\bf r})\right]\, , \ea
and for the gravitational potential $\varphi$ and the radiation fluctuations $\delta\varepsilon_{\mathrm{rad2}}$ we obtain the system
of equations:
\ba{3.22}
\triangle\varphi+3\mathcal{K}\varphi&=&\frac{\kappa c^4}{2}\delta\rho-2\kappa c^2A_0f({\bf r})\, ,\\ \label{3.23}
\delta\varepsilon_{\mathrm{rad2}}&=&-12A_0\frac{\varphi}{a^4c^2}\, .
\ea
Similar to the Model I, here we also get the same physically reasonable results.


\section{Flat rotation curves}

\setcounter{equation}{0}

It is well known that rotation curves of disc galaxies have the flat shape starting from  some distance. The real reason of such shape is still unclear. To
explain it, different mechanisms were proposed from Modified Newtonian Dynamics and other modifications of gravity (see, e.g., \cite{Mofat0}) to the presence
of dark matter or other specific fields. For example, the nonrelativistic gravitational potential in a galaxy may be presented as follows
\cite{Sanders,Moffat}:
\ba{4.1} \varphi_\mathrm{ph}({\bf R})&=&-\frac{G_NM}{R}[1+\alpha\exp(-R/R_0)]\nn\\
&=&-\frac{G_NM}{R}-\frac{G_NM}{R}\alpha\exp(-R/R_0)\, . \ea
Here, $\varphi_\mathrm{ph}$ is the physical (not comoving) potential and $R$ is the physical distance from the center of a galaxy\footnote{We have mentioned in
section 3 that the physical distance $R$ and the comoving distance $r$ are connected as follows: $R=ar$. Obviously, there is no need to take into account the
dynamics of the Universe in the case of astrophysical problems, i.e. here the scale factor $a$ is considered as a constant value.}. $R_0$ is the Yukawa
interaction range, $\alpha$ is the coupling strength and $M$ is the total effective mass at infinity.  To get the flat rotation curves, the additional Yukawa
term must result in a repulsive force, i.e. $\alpha <0$.

The Eqs. \rf{3.19} and \rf{3.22} clearly indicate that the QN distribution affect the gravitational potential. Can we get the potential of the form \rf{4.1},
which is motivated by the observational data, from these equations\footnote{Clearly, we can consider other forms of the potential and find for them
corresponding QN distributions. The only restriction here is the demand that such potentials provide the rotation curves in accordance with observations.}? In
other words, what kind of the distribution function $f(\bf r)$ should be used to provide \rf{4.1}? To answer this question, we rewrite Eqs. \rf{3.19} and
\rf{3.22} in the astrophysical setting. This means that we put $\mathcal{K}=0$, $\delta\rho=\rho$ and consider physical values instead of comoving. Then, Eq.
\rf{3.19} reads
\be{4.2}
\triangle_R\varphi_\mathrm{ph}=4\pi G_N\rho_\mathrm{ph}+4\pi G_N\frac{3(-A_1)^{4/3}}{c^2(A_4)^{1/3}}f_\mathrm{ph}({\bf R})\, ,
\ee
where $\varphi_\mathrm{ph}$, $\rho_\mathrm{ph}$ and $f_\mathrm{ph}({\bf R})$ are physical values and the Laplace operator $\triangle_R$ is defined with respect
to the physical distance $\bf R$. To get this equation, we divide both sides of \rf{3.19} by $a^3$. For example, $\varphi_\mathrm{ph}=\varphi/a$,
$\rho_\mathrm{ph}=\rho/a^3$ and $f_\mathrm{ph}({\bf R})= f({\bf r})/a^3$. As we wrote in the footnote 4, we neglect the time dependence of the scale factor $a$
in the astrophysical setting.

Let $\rho_\mathrm{ph}$ describe the rest mass density of the pure baryonic matter. We simulate it in the delta-shape form: $\rho_\mathrm{ph}=m\delta({\bf R})$,
where $m$ is the mass of the baryonic constituent. Then, the substitution of the potential \rf{4.1} into Eq. \rf{4.2} leads to the following function
$f_\mathrm{ph}(\bf R)$:
\be{4.3}
f_\mathrm{ph}({\bf R})=-\frac{M\alpha c^2}{12\pi RR_0^2}\frac{(A_4)^{1/3}}{(-A_1)^{4/3}}\exp(-R/R_0)\, ,
\ee
which describes the QN distribution.

Similarly, in the case of Model II Eq. \rf{3.22} reads
\be{4.4}
\triangle_R\varphi_\mathrm{ph}=4\pi G_N\rho_\mathrm{ph}-16\pi G_N \frac{A_0}{c^2}f_\mathrm{ph}({\bf R})
\ee
and the required distribution of QNs has the form
\be{4.5}
f_\mathrm{ph}({\bf R})=\frac{M\alpha c^2}{16\pi RR_0^2}\frac{1}{A_0}\exp(-R/R_0)\, .
\ee
For both of these models the effective mass $M$ and the bare baryonic mass $m$ are related as follows: $M(1+\alpha)=m$.

It makes sense to rewrite the distribution functions \rf{4.3} and \rf{4.5} via the parameter $\gamma$ which was estimated for some cosmological models in
\cite{BEJZ1} and was also restricted experimentally in \cite{Bilbao}. The most simple case corresponds to the model where QNs are the only possible
representatives of dark matter (this is the $\beta=0$ case in these papers). Here, we have the pure $\Lambda$CDM model with clear origin of dark matter.
According to \cite{BEJZ1}, the parameter $\gamma$ is determined as follows
\ba{4.6}
\frac{8\pi G_N}{c^2}\frac{(-A_1)^{4/3}}{(A_4)^{1/3}}&=&\gamma^{4/3}a_0^3H_0^2\, ,\\
\label{4.7}
-\frac{8\pi G_N}{c^2}A_0&=& \frac{3}{4}\gamma a_0^3H_0^2
\ea
for Models I and  II, respectively. Here, $a_0$ and $H_0$ are the scale factor and the Hubble parameter, respectively, at the present moment. Then, the QN
distribution functions take the form
\be{4.8} f_\mathrm{ph}({\bf R})= -\frac{2}{3}\, \frac{M\alpha G_N}{\gamma^{4/3}a_0^3H_0^2}\, \frac{1}{RR_0^2}\, e^{- R/R_0}\, ,\ \mbox{Model I} \ee
and
\be{4.9} f_\mathrm{ph}({\bf R})= -\frac{2}{3}\, \frac{M\alpha G_N}{\gamma a_0^3H_0^2}\, \frac{1}{RR_0^2}\, e^{- R/R_0}\, ,\ \mbox{Model II}\, . \ee
Taking into account the inequalities $\gamma>0$ and $\alpha<0$, we see that these functions describe the overdensities. This is the physically reasonable
result. In addition, we would like to stress that similar profile functions are really used in literature for resolving the rotation curves flatness problem
(see, e.g., the Prugniel-Simien model discussion in \cite{PSmodel}). Besides, it is worth mentioning that to solve this problem, in \cite{Rahaman} the authors
also investigated (in a different manner) the quark-gluon plasma as dark matter in the halos of galaxies.


\section{Conclusion}

In our paper, we have studied the Universe filled with the dustlike matter (baryonic and dark), radiation and quark-gluon nuggets. The Universe has been
considered at late stages of its evolution and at scales much less than the cell of uniformity size which is approximately 190 Mpc \cite{EZcosm2}. At such
distances, our Universe is highly inhomogeneous and the averaged Friedmann approach does not work here. We need to take into account the inhomogeneities in the
form of galaxies, groups and clusters of galaxies. It is natural to assume also that radiation as well as quark-gluon nuggets fluctuate around the average
values. Therefore, these fluctuations as well as inhomogeneities perturb the FRW metrics. To consider these perturbations inside the cell of uniformity, we
need to use the mechanical approach. This approach was established in our papers \cite{EZcosm1,EKZ2,EZcosm2}. An important feature of this approach is that it
provides an opportunity to study self-consistency of different cosmological models (see, e.g., \cite{BUZ1}). For example, there is a possibility that a small
fraction of colored objects escaped hadronization and survived in the form of quark-gluon nuggets \cite{Witten}. Therefore, it is of interest to investigate
the compatibility of such QNs with the scalar perturbations theory. This was the main aim of our studies.

We have considered two models which have different equations of state. For both of these models, we got similar results which look physically reasonable.
First, the nonrelativistic gravitational potential is defined by the distribution of inhomogeneities/fluctuations of both dustlike matter and QNs (see the
corresponding equations \rf{3.19} and \rf{3.22}). To find the exact form of the potential, we need to know the distribution of dustlike inhomogeneities (i.e.
the function $\delta \rho(\bf r)$ which is the difference between the real and averaged rest mass densities) and the distribution of fluctuations of QNs (i.e.
the function $f(\bf r)$). Therefore, the nonrelativistic gravitational potential is determined by the distribution of both the baryonic inhomogeneities and
quark-gluon nuggets. Consequently, we demonstrated that QNs can be distributed around baryonic inhomogeneities (e.g., galaxies) in such a way that it can solve
the problem of the flatness of the rotation curves. Therefore, flat rotation curves can be explained with the help of particles from the standard model of high
energy physics, i.e. without involvement of exotic particles or modification of gravity. This is an advantage of our approach.
Second, the fluctuations of radiation are caused by both the inhomogeneities in the form of galaxies (see Eq. \rf{3.8}) and the fluctuations of quark-gluon
nuggets (see Eqs. \rf{3.20} and \rf{3.23}). Therefore, if QNs exist, the CMB anisotropy contains also the contributions from QNs. Additionally, the spatial
distribution of the radiation fluctuations is defined by the gravitational potential $\varphi (\bf r)$ that is also quite reasonable. On the whole, our study
showed that quark-gluon nuggets can be compatible with the mechanical approach. The authors of the paper \cite{Bilbao} also found that our models can be in
agreement with the recent experimental data.


\section*{Acknowledgements}

The work of M. Eingorn was supported by NSF CREST award HRD-1345219 and NASA grant NNX09AV07A.




\begin{thebibliography}{99}
\bibitem{LasloYad1}
L.L. Jenkovszky, B. K\"ampfer and V.M. Sysoev, Z. Phys. C, Particles and Fields {\bf 48} (1990) 147.
\bibitem{LasloYad2}
V.G. Boyko, L.L. Jenkovszky, B. K\"ampfer and V.M. Sysoev, J. Nucl. Phys. {\bf 51} (1990) 1134.
\bibitem{Tillmann} T. Boeckel and J. Schaffner-Bielich, Phys. Rev. D {\bf 85} (2012) 103506; arXiv:astro-ph/1105.0832.
\bibitem{Witten}
E. Witten, Phys. Rev. D {\bf 30} (1984) 272.
\bibitem{Applegate}
A. Applegate and C.J. Hogan, Phys. Rev. D {\bf 31} (1985) 3037.
\bibitem{Farhi}
E. Farhi and R.L. Jaffe, Phys. Rev. D {\bf 30} (1984) 2379.
\bibitem{Chandra}
D. Chandra and A. Goyal, Phys. Rev. D {\bf 62} (2000) 063505; arXiv:hep-ph/9903466.
\bibitem{BEJZ1}
M. Brilenkov, M. Eingorn, L. Jenkovszky and A. Zhuk, JCAP {\bf 08} (2013) 002; arXiv:astro-ph/1304.7521.
\bibitem{Bhatta}
A. Bhattacharyya et al., Nucl. Phys. A {\bf 661} (1999) 629; arXiv:hep-ph/9907262.
\bibitem{Bhatta2}
A. Bhattacharyya et al., Phys. Rev. D {\bf 61} (2000) 083509; arXiv:hep-ph/9901308.
\bibitem{BASR}
P. Bhattacharjee, J. Alam, B. Sinha and S. Raha, Phys. Rev. D {\bf 48} (1993) 4630.
\bibitem{ARS}
J. Alam, S. Raha and B. Sinha, Astrophys J. {\bf 513} (1999) 572; arXiv:astro-ph/9704226.
\bibitem{Gh}
S. Ghosh, {\it Astrophysics of Strange Matter}, Plenary Talk at 2008 Quark Matter, Jaipur, India; arXiv:astro-ph/0807.0684.
\bibitem{Kalam}
M. Kalam et al., Int. J. Theor. Phys. {\bf 52} (2013) 3319; arXiv:gr-qc/1205.6795.
\bibitem{Zhitnitsky}
A.R. Zhitnitsky, JCAP {\bf 10} (2003) 010; arXiv:hep-ph/0202161.
\bibitem{Perez}
M.A. Perez-Garcia, J. Silk and J.R. Stone, Phys. Rev. Lett. {\bf 105} (2010) 141101; arXiv:astro-ph/1007.1421.
\bibitem{Drake}
J.J. Drake et al.,
Astrophys. J. {\bf 572} (2002) 996; arXiv:astro-ph/0204159.
\bibitem{Madsen}
J. Madsen, Lect. Notes Phys. {\bf 516} (1999) 162; arXiv:astro-ph/9809032.
\bibitem{Madsen2}
J. Madsen and J.M. Larsen, Phys. Rev. Lett. {\bf 90} (2003) 121102; arXiv:astro-ph/0211597.
\bibitem{Madsen3}
J. Madsen, Phys. Rev. D {\bf 71} (2005) 014026; arXiv:astro-ph/0411538.
\bibitem{EZcosm1}
M. Eingorn and A. Zhuk, JCAP {\bf 09} (2012) 026; arXiv:astro-ph/1205.2384.
\bibitem{EKZ2}
M. Eingorn, A. Kudinova and A. Zhuk, JCAP {\bf 04} (2013) 010; arXiv:astro-ph/1211.4045.
\bibitem{EZcosm2}
M. Eingorn and A. Zhuk,  JCAP {\bf 05} (2014) 024; arXiv:astro-ph/1309.4924.
\bibitem{BUZ1}
A. Burgazli, M. Eingorn and A. Zhuk, {\em Rigorous theoretical constraint on constant negative EoS parameter $\omega$ and its effect for the late Universe};
arXiv:astro-ph/1301.0418.
\bibitem{Bilbao}
A. Montiel, V. Salzano and R. Lazkoz,  Phys. Lett. B {\bf 733C} (2014) 209;
arXiv:astro-ph/1404.0388.
\bibitem{Kallmann}
C.G. K\"allmann, Phys. Lett. B {\bf 134} (1984) 363.
\bibitem{LasloEch}
V.G. Boyko, L.L. Jenkovszky and V.M. Sysoev, EChAYa {\bf 22} (1991) 675.
\bibitem{Begun2004}
V.V. Begun, M.I. Gorenstein and O.A. Mogilevsky, Int. J. Mod. Phys. E {\bf 20} (2011) 1805; arXiv:hep-ph/1004.0953.
\bibitem{Pisarski2006}
R.D. Pisarski, Phys. Rev. D {\bf 74} (2006) 121703.
\bibitem{Pisarski2007}
R.D. Pisarski, Progr. Theor. Phys. Suppl. {\bf 168} (2007) 276.
\bibitem{Landau}
L.D. Landau and E.M. Lifshitz, {\em The Classical Theory of Fields, Fourth Edition: Volume 2 (Course of Theoretical Physics Series)}, Oxford Pergamon Press,
Oxford (2000).
\bibitem{Mukhanov}
V.F. Mukhanov, H.A. Feldman and R.H. Brandenberger, Physics Reports {\bf 215} (1992) 203.
\bibitem{Rubakov}
D.S. Gorbunov and V.A. Rubakov, {\em Introduction to the Theory of the Early Universe: Cosmological Perturbations and Inflationary Theory}, World Scientific,
Singapore (2011).
\bibitem{Mofat0}
J.W. Moffat, JCAP {\bf 0603} (2006) 004; arXiv:gr-qc/0506021.
\bibitem{Sanders}
R.H. Sanders, Astron. Astrophys. {\bf 136} (1984) L21; Astron. Astrophys. {\bf 154} (1986) 135.
\bibitem{Moffat}
J.W. Moffat, JCAP {\bf 0505} (2005) 003; arXiv:astro-ph/0412195.
\bibitem{PSmodel}
D. Merritt, A.W. Graham, B. Moore, J. Diemand and B. Terzic, Astron. J. {\bf 132} (2006) 2685; arXiv:astro-ph/0509417; Astron. J. {\bf 132} (2006) 2701;
arXiv:astro-ph/0608613.
\bibitem{Rahaman}
F. Rahaman, P.K.F. Kuhfittig, R. Amin, G. Mandal, S. Ray and N. Islam, Phys. Lett. B {\bf 714} (2012) 131; arXiv:gr-qc/1203.6649.

\end{thebibliography}
\end{document}